\documentclass[lettersize,journal]{IEEEtran}
\usepackage{amsmath,amsfonts}
\usepackage{algorithmic}
\usepackage{algorithm}
\usepackage{array}
\usepackage[utf8]{inputenc}
\usepackage{subcaption} 
\usepackage{cleveref} 

\usepackage{stfloats} 
\usepackage{xcolor}
\usepackage{textcomp}
\usepackage{url}
\usepackage{verbatim}
\usepackage{graphicx}
\usepackage{cite}
\usepackage{svg}
\begin{document}

\title{Generative-AI for XR Content Transmission in the Metaverse: Potential Approaches, Challenges, and a Generation-Driven Transmission Framework}

\author{Zhe Zhang,~\IEEEmembership{Senior Member,~IEEE,}
        Yili~Jiang,~\IEEEmembership{Member,~IEEE,}
        Xin~Wei,~\IEEEmembership{Senior Member,~IEEE,}
        Mingkai~Chen,~\IEEEmembership{Member,~IEEE,}
        Haiwei~Dong,~\IEEEmembership{Senior Member,~IEEE,}
        Shui~Yu,~\IEEEmembership{Fellow,~IEEE}
        
}


\maketitle


\section*{Abstract}
How to efficiently transmit large volumes of Extended Reality (XR) content through current networks has been a major bottleneck in realizing the Metaverse. The recently emerging Generative Artificial Intelligence (GAI) has already revolutionized various technological fields and provides promising solutions to this challenge. In this article, we first demonstrate current networks' bottlenecks for supporting XR content transmission in the Metaverse. Then, we explore the potential approaches and challenges of utilizing GAI to overcome these bottlenecks. To address these challenges, we propose a GAI-based XR content transmission framework which leverages a cloud-edge collaboration architecture. The cloud servers are responsible for storing and rendering the original XR content, while edge servers utilize GAI models to generate essential parts of XR content (e.g., subsequent frames, selected objects, etc.) when network resources are insufficient to transmit them. A Deep Reinforcement Learning (DRL)-based decision module is proposed to solve the decision-making problems. Our case study demonstrates that the proposed GAI-based transmission framework achieves a 2.8-fold increase in normal frame ratio (percentage of frames that meet the quality and latency requirements for XR content transmission) over baseline approaches, underscoring the potential of GAI models to facilitate XR content transmission in the Metaverse.

\section{Introduction}
The concept of the Metaverse was originally envisaged by Neal Stephenson in his science fiction novel ``Snow Crash" in 1992. Driven by rapid technological advancements, such as Digital Twin, Artificial Intelligence (AI), and Extended Reality (XR) which includes  Virtual Reality (VR) and Augmented Reality (AR), Metaverse has evolved from speculative fiction into a tangible reality. Moreover, the introduction of haptic feedback not only enhances the realism of virtual interactions, but also paves the way for multi-sensory services in the Metaverse.


The ambitious goal of Metaverse is to become the evolution of the Internet that can create a truly immersive and interactive virtual world where billions of people can seamlessly engage in work, entertainment, socialization, etc. However, existing approaches relying on end devices like headsets for rendering XR content face limitations due to their computational power. Hence, leveraging the computational capabilities of cloud or edge servers offers a promising solution \cite{Haiwei}. Yet, relying solely on edge servers for XR rendering is unfeasible due to resource constraints, while cloud servers introduce long-distance latency, which is unacceptable for real-time interactive applications. Moreover, there is currently no simple method to combine the strengths of edge and cloud computing for efficient XR content transmission \cite{A-Full-Dive}. 

The recent emerging Generative Artificial Intelligence (GAI) has profoundly reshaped the future of technology in different aspects. By generating complex content at edge servers rather than transmitting it from remote cloud servers, GAI minimizes the amount of data that must be transmitted across the network. This reduces traffic load and significantly enhances transmission efficiency. 
To leverage GAI for XR content transmission in the Metaverse, we propose a GAI-based framework with a cloud-edge collaboration architecture. The original XR content is stored and rendered in the cloud, while GAI models at edge servers generate essential parts of the content when network resources are insufficient. This approach guarantees smoother XR content delivery by generating and transmitting crucial parts (e.g., subsequent frames, objects, and background scenes) from edge servers to users when link bandwidth is insufficient. Our framework offers a scalable solution for reliable XR services, making it suitable for diverse Metaverse applications. The main contributions of this paper are as follows:
\begin{itemize}
    \item We discuss the challenges of XR content transmission in the Metaverse, including the need for high-quality video content and haptic feedback, and explore how GAI can address these challenges while discussing the associated implementation issues.
    \item A GAI-based XR content transmission framework is proposed to realize the potential approaches of utilizing GAI for XR content (both video and haptic) transmission and address some of the related challenges.
    \item We introduce a Deep Reinforcement Learning (DRL)-based decision module to overcome the remaining challenges, particularly the decision-making related challenges.
    \item We evaluate the subsequent frame generation module through a case study, demonstrating that our approach achieves higher normal frame ratio (frames meeting XR quality and latency requirements) and lower end-to-end latency than baseline approaches.
\end{itemize}

The paper is organized as follows: We first analyze the network challenges in Metaverse content transmission, then explore the potential and challenges of using GAI to improve XR transmission. Next, we present our GAI-based approach, followed by a case study to demonstrate its effectiveness. Finally, we conclude with future research directions.




\section{Network Challenges for Supporting Content Transmission in the Metaverse}
We discuss current network challenges for supporting XR content transmission in the Metaverse in detail in this section.



\subsection{Intelligent Adaptive Transmission for Large Volume of XR Content}
Efficient transmission of high-resolution XR content is critical for immersive experiences. Unlike regular 2D videos, Metaverse multimedia formats (e.g., light field videos, stereo videos, volumetric videos, and meshes) demand significantly more data. For instance, volumetric videos, enabling fully interactive 3D experiences, require data rates up to 2.9 Gb/s \cite{Volumetric-Video}. Consequently, current networks cannot transmit all XR content at full quality. Prioritization based on user behavior, content context, Regions of Interest (ROI), and network conditions is essential, requiring intelligent mechanisms to identify key content and allocate resources adaptively in real time.

\subsection{Seamless Interaction}

Seamless interaction between users and virtual environments is crucial for the Metaverse, particularly with haptic feedback (e.g., Metaverse surgery), which requires a Round-Trip Time (RTT) of 1 ms \cite{tactile}. Achieving this low latency is challenging for current networks, especially from remote cloud servers. Prediction technologies, such as viewpoint prediction \cite{prediction}, offer a solution by pre-rendering content outside the user’s view. They also enhance multi-modal interactions by predicting potential XR content like haptic feedback based on user behavior, ensuring seamless integration of touch and movement with the virtual environment.

\subsection{Consistently Stable Transmission in Dynamic Network for XR Content}

XR content requires stable and reliable network connections due to its high data rates and real-time needs. However, dynamic network conditions, including fluctuating bandwidth, latency, and packet loss, especially in wireless environments \cite{network}. Unlike traditional content, where buffering can mitigate issues like frame drops, XR content demands uninterrupted, stable connections, necessitating innovative solutions to maintain stable transmission in dynamic conditions.



\subsection{Challenges of Ubiquitous Access and Device Heterogeneity in XR Transmission}

XR content is processed by diverse devices, from AR glasses and VR headsets to haptic gloves \cite{A-Full-Dive}. To ensure an immersive experience, the network must support ubiquitous access for these devices, such as using Wi-Fi indoors and 5G outdoors. This requires handling the varying computational capabilities of each device and adapting to different network technologies and performance needs. Conventional approaches are insufficient for such diverse requirements, highlighting the need for intelligent scaling and optimization for both devices and networks.



\section{Potential Approaches and Challenges of Utilizing GAI for XR Content Transmission}

GAI offers promising solutions to the network challenges discussed earlier. This section reviews popular GAI models, explores their potential for enhancing XR content transmission in the Metaverse, and discusses the challenges of implementing GAI for XR content delivery.

\subsection{GAI Models}

\begin{table*}[htp]
\centering
\renewcommand{\arraystretch}{1.5} 
\small 
\caption{Summary of GAI Models}
\begin{tabular}{|>{\centering\arraybackslash}m{2.5cm}|>{\centering\arraybackslash}m{3.5cm}|>{\centering\arraybackslash}m{4.0cm}|>{\centering\arraybackslash}m{3.0cm}|>{\centering\arraybackslash}m{3.0cm}|}
\hline
\textbf{GAI Model} & \textbf{Key Characteristics} & \textbf{Advantages} & \textbf{Limitations} & \textbf{Applications}   \\ \hline

Generative Adversarial Networks (GANs) \cite{gan} & 
Adversarial training, \newline Generator-discriminator framework &
No need for extensively labeled training data, \newline Less inference time, &
Unstable training, \newline Computationally intensive &
Text generation, \newline Image generation, \newline Video generation  
\\ \hline

Transformer Models \cite{attention} & 
Self-attention mechanism, \newline Encoder-decoder architecture &
Sequential data dependency capture capability, \newline Parallelized training, \newline Model integration capability&
Computationally intensive &
Natural Language Processing (NLP), \newline Translation, \newline Image generation 
\\ \hline

Diffusion Models (DMs) \cite{dm} & 
Incremental noise introduction, \newline Denoising process &
More stable training, \newline Realistic content generation &
Computationally intensive &
Text generation, \newline Image generation, \newline Video generation  
\\ \hline

Large Generative AI Models (LGAIMs) \cite{scaling} & 
Billions of parameters &
Capable of generating more complex and diverse content &
Tremendous resource requirement for training and inference &
Text generation, \newline Image generation, \newline Video generation  
\\ \hline
\end{tabular}
\label{GAI model} 
\end{table*}

We briefly review several popular GAI models, including GANs \cite{gan}, Transformer Models \cite{attention}, Diffusion Models (DMs) \cite{dm}, and LGAIMs \cite{scaling}, which are essential for addressing XR content transmission challenges, particularly in content generation. Selecting the appropriate GAI model depends on task requirements, resources, and application needs. Key characteristics, advantages, limitations, and typical applications are summarized in Table \ref{GAI model}.

\subsection{Potential Directions of Utilizing GAI for XR Content Transmission}
\begin{figure*}[!htp]
\centering {\includegraphics[width=1\textwidth]{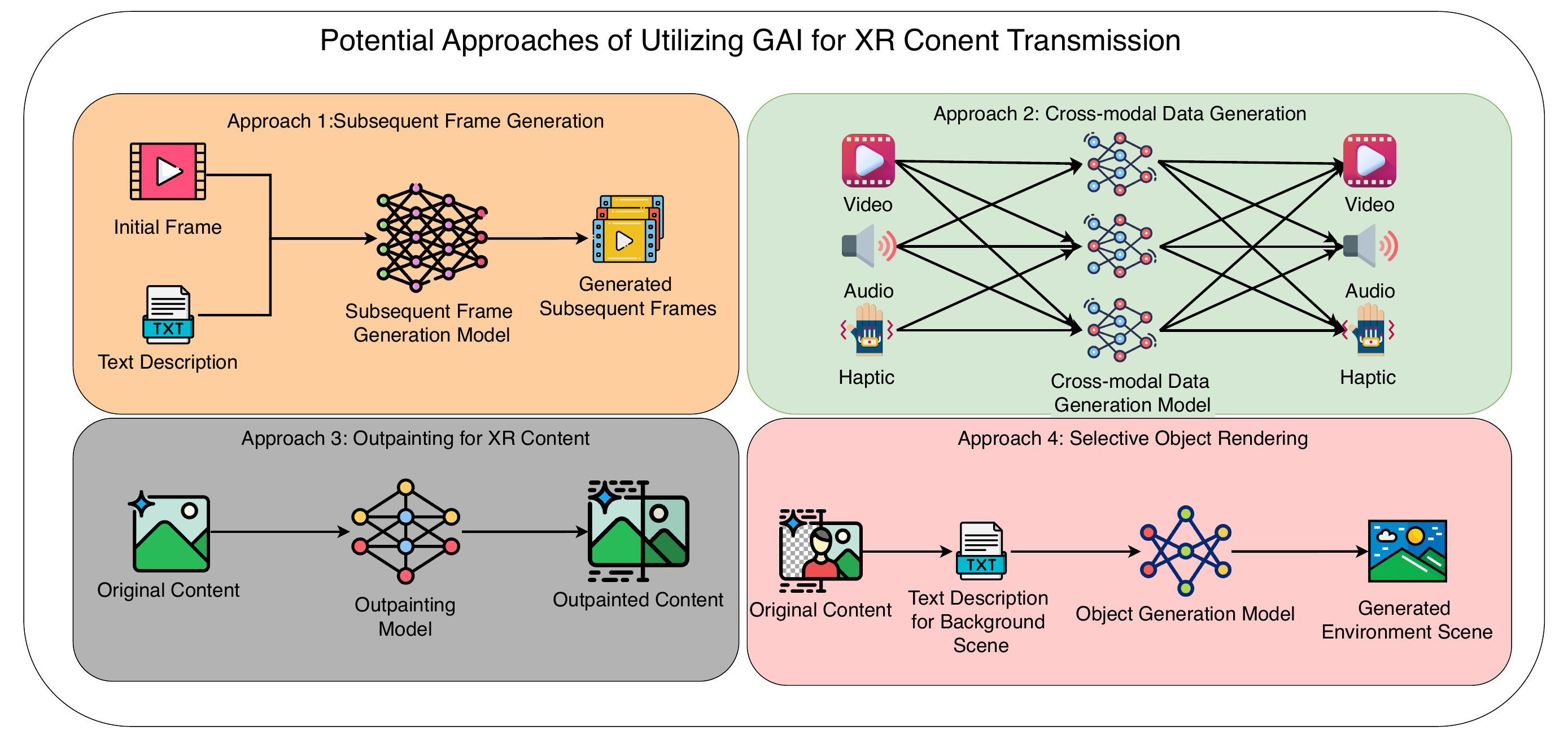}}
\caption{The potential approaches to leverage GAI to facilitate XR content transmission in the Metaverse.} 
\label{app} 
\end{figure*}

While extensive research has explored the impact of GAI on Metaverse content, its role in facilitating XR content transmission remains underexplored. This subsection addresses this gap by discussing potential approaches for leveraging GAI to enhance XR content transmission, as illustrated in Fig. \ref{app}. These four approaches can be used independently or jointly, depending on the application's requirements and network condtitions.

\subsubsection{Subsequent Frame Generation}
To guarantee users' immersive experience, all XR content frames must be rendered and transmitted in real-time. However, achieving this goal is challenging due to unpredictable traffic, unstable connections, and long-distance latency.

One possible solution is to leverage GAI to generate subsequent frames at edge servers when network conditions fail to satisfy the transmission requirements. By transmitting only text descriptions or base frames from cloud servers, GAI models at the edge can generate high-quality XR frames, significantly reducing bandwidth usage. This ensures continuous and immersive Metaverse experiences despite network fluctuations.


\subsubsection{Cross-modal Data Generation}
Multi-modal data, such as video, audio, and haptic feedback, is essential for the Metaverse to provide immersive experiences to users. However, each modality of data has unique QoS requirements, making it challenging to guarantee consistent service across all modalities simultaneously. For example, haptic feedback requires extremely low latency (usually about 1 ms for RTT), which is difficult to satisfy when transmitted from cloud servers due to the multi-hop transmission.


A promising solution is leveraging GAI for cross-modal data generation, such as predicting haptic feedback from video. For example, texture and force feedback can be predicted from a video of a user interacting with a plastic ball. By pre-generating haptic feedback, it can be delivered instantly from edge servers, meeting latency requirements. However, challenges include maintaining spatial and temporal coherence, as prediction inaccuracies may lead to mismatches. Synchronization can be improved by adding timestamps in packet headers \cite{wei}.


\subsubsection{Outpainting for XR content}
In XR environments, users interact with expansive virtual worlds that extend beyond their immediate view. Rendering these extended areas requires collecting user behavior information and transmitting it back to the cloud or edge servers for rendering, which introduces additional latency and degrades the seamless experience expected in the Metaverse.

GAI enables cloud or edge servers to preemptively outpaint extended areas of the virtual environment based on the current scene, eliminating the need for detailed user behavior data \cite{network}. This allows the generated areas to keep pace with user movements with minimal latency, ensuring a continuous and immersive experience. The key challenge lies in accurately predicting and outpainting the areas that the user will view next.



\subsubsection{Selective Object Rendering}
Usually, XR users only care about objects within their Field of View (FoV). Hence, rendering and transmitting every object within the scene is resource-intensive and demands significant bandwidth, potentially affecting the real-time performance of the application.

Typically, objects within the users’ FoV are prioritized for high-quality rendering, while other objects can be treated as less important \cite{fov}. To optimize resource allocation, the cloud transmits high-quality rendered key objects and text descriptions for less important objects. Although generated objects may lack perfect fidelity, their reduced quality minimally impacts user experience, as they aren't the focus of attention. Notably, the selective object rendering also applies in 6 Degrees of Freedom (6DoF) since users in 6DoF environments are also focused on certain objects.



\subsection{Challenge of Integrating GAI for XR Content Transmission}
Although utilizing GAI for XR content transmission brings promising solutions to overcome current network bottlenecks, several significant challenges have to be addressed to fully realize its potential. We explore the key challenges in detail as follows:

\subsubsection{High Computational Demands Challenge}
Generating high-quality XR content requires substantial computational resources, which increase with content quality and length. Compared to cloud servers, edge servers usually have limited computational resources. Thus, how to effectively balance and allocate resources among various tasks will be a crucial challenge. 

\subsubsection{Long Inference Latency Challenge}
The inference latency of GAI models, which refers to the time required to generate XR content, is usually significant. However,real-time Metaverse applications, like gaming and surgery, have strict latency requirements. Even minor delays can significantly degrade user experience. Thus, the long inference latency is unacceptable for those applications. How to mitigate the impacts of the long inference latency of GAI models is a must-solve challenge. 


\subsubsection{Decision-Making for Content Generation Over Dynamic Networks Challenge}
Multiple decision-making challenges need to be addressed in using GAI for content generation. These include determining how the cloud servers and edge servers cooperate, deciding the quality and length of the generated XR content, and identifying the optimal timing for activating GAI models. Additionally, these decisions are significantly impacted by dynamic networks. As a result, making appropriate decisions for content generation over dynamic networks is another crucial challenge.

\subsubsection{Prediction Accuracy}

Prediction accuracy refers to the similarity between generated and original content, which is affected by the probabilistic nature of GAI models. These models generate content based on learned patterns, but inherent randomness often causes variations (e.g., in textures or colors), leading to inconsistencies. If these differences are too significant, user experience can degrade. Furthermore, prediction accuracy is crucial for spatial coherence in cross-modal data generation. Low accuracy may cause mismatches between modalities, affecting the immersive experience. Mitigating the impact of these variations to improve accuracy remains a key challenge \cite{network}.

\subsubsection{Security and Privacy Challenge}
The inputs for GAI model inference could be the XR content itself, descriptions of this content, or even user behavior information. These inputs could expose sensitive information such as user biometric information, personal preferences, behavior patterns, etc. When cloud or edge servers utilize GAI models, they can have access to this sensitive data, which raises the risk of unauthorized collection, analysis, or misuse of user information \cite{zhou2024survey}. Therefore, multiple techniques, including encryption, anonymization, and strict access control, are needed to protect user privacy in the Metaverse.


\section{GAI-based Content Transmission Framework}

\begin{figure*}[!htp]
\centering {\includegraphics[width=1\textwidth]{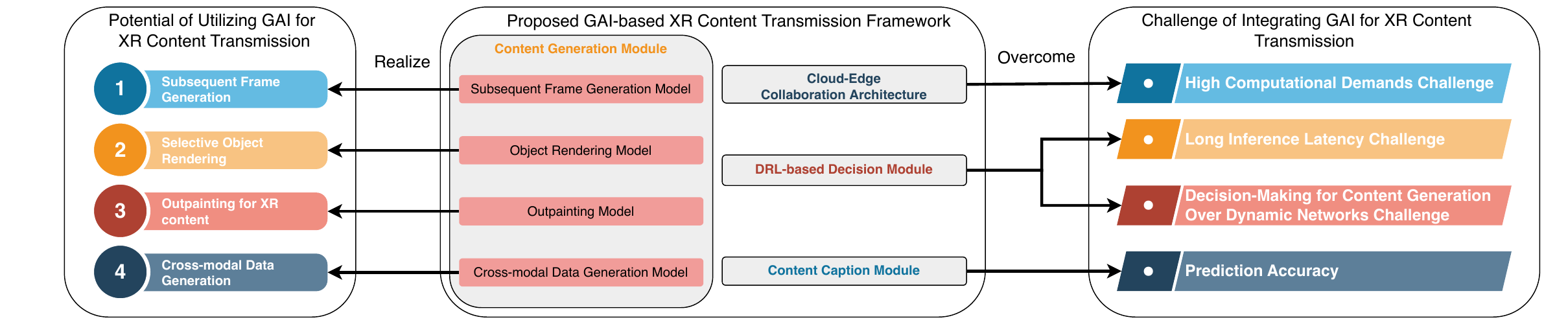}}
\caption{The relationship between the potential approaches, related challenges, and the proposed GAI-based transmission framework.} 
\label{flow} 
\end{figure*}

To leverage the benefits of GAI and address the challenges outlined in the previous section, we propose a GAI-based XR content transmission framework. Fig. \ref{flow} shows the relationship between the potential approaches, associated challenges, and the proposed GAI-based transmission framework.

\begin{figure*}[!htp]
\centering {\includegraphics[width=1\textwidth]{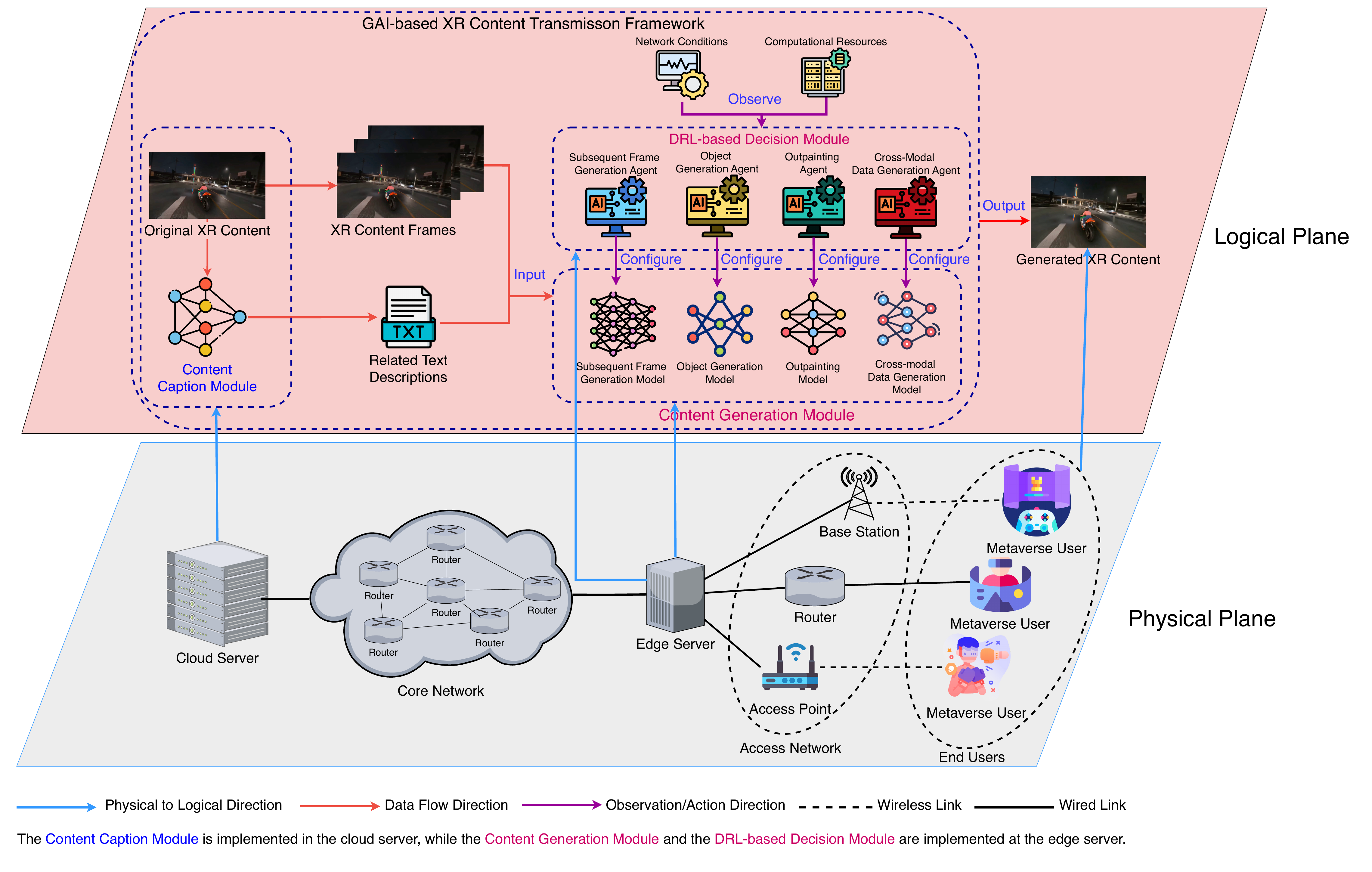}}
\caption{The workflow of the proposed GAI-based content transmission framework.} 
\label{proposed} 
\end{figure*}

The proposed framework leverages a cloud-edge collaboration architecture and consists of three modules: a content caption module, a content generation module, and a DRL-based decision module. Fig. \ref{proposed} illustrates the workflow of the proposed GAI-based content transmission framework.

\subsection{Cloud-Edge Collaboration Architecture}
\begin{figure}[!htp]
\centering {\includegraphics[width=0.4\textwidth]{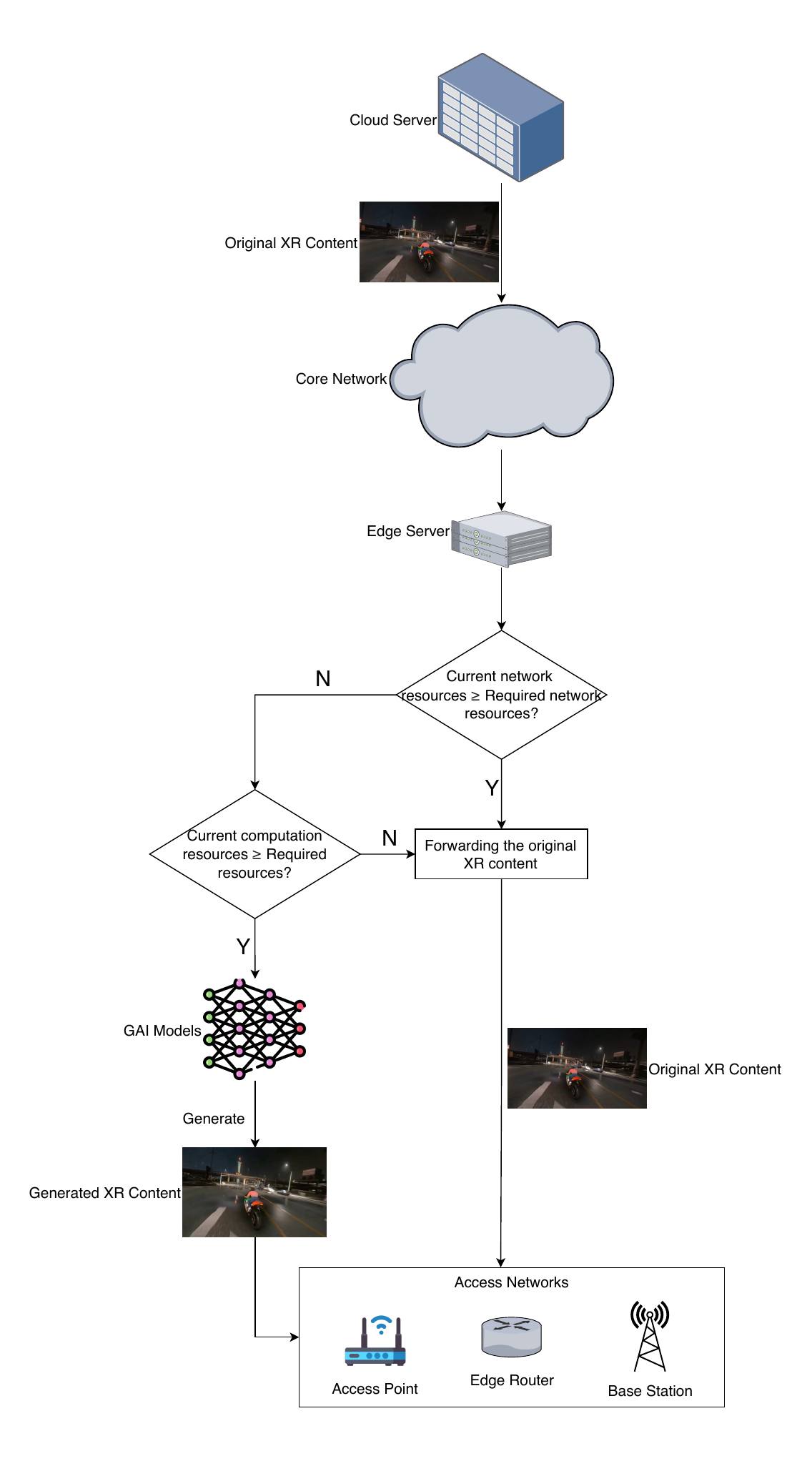}}
\caption{The interaction between cloud servers and edge servers.} 
\label{cloud-edge} 
\end{figure}

To address the high computational demands and long inference latency challenges that are mentioned in the previous section, we adopt a cloud-edge collaboration architecture. The cloud servers and edge servers collaborate to decide the task assignments based on the current network conditions, edge servers' available computational resources, and the QoS requirements of XR content. Specifically, the original XR content will be directly transmitted from the cloud servers to end users if the network conditions can satisfy the XR content's QoS requirements. Otherwise, the edge servers will leverage GAI models to generate essential parts of the XR content when they have enough computational resources. The collaboration between cloud servers and edge servers is shown in Fig. \ref{cloud-edge}.

\subsection{Content Generation Module}
The content generation module plays a key role in the proposed GAI-based content transmission framework. It is implemented on the edge servers and can generate essential content (such as subsequent frames, selected objects, and background scenes) using various state-of-the-art GAI models when network conditions are inefficient to transmit the original XR content from the remote cloud servers. For instance, GANs can be used to generate less detailed content, e.g., background objects in an XR scene, while Diffusion Models (DMs) can be used to generate highly realistic content, e.g., objects that users can directly interact with \cite{zhou2024survey}. The content generation module consists of:
\begin{itemize}
\item \textbf{A Subsequent Frame Generation Model}: It can generate subsequent frames based on related text descriptions or based frames by utilizing GAI models. For example, both DMs and GANs can be used to generate high-quality XR frames. GANs are preferable for low-latency scenarios as they do not require iterative refinement, offering faster performance. On the other hand, DMs provide higher content quality but come with increased processing time due to their iterative nature. Therefore, even when network conditions are not sufficient to transmit original XR content from remote cloud servers, the immersive experience of XR users can still be maintained with the generated frames.

\item \textbf{A Cross-modal Data Generation Model}: GAI models can generate one modality of content from another, reducing the need to transmit all modalities. This is particularly useful when network resources are limited, as it ensures continuity in the XR experience and enables seamless user interaction. Due to the capability of processing varying length sequences, transformer-based models are suitable for generating one modal content from other modalities. 
\item \textbf{An Object Rendering Model}: This model uses GAI to generate less important objects (typically outside the ROI) that should originally be rendered by cloud servers to save computational resources and ensure the real-time performance of XR applications. For example, GANs can be used to generate background scenes since they can quickly generate less detailed content, which is sufficient for areas outside the ROI.
\item \textbf{An Outpainting Model}: This model can preemptively conduct outpainting for extended areas by leveraging GAI models in advance. The cloud servers can only transmit content within the ROI, which can significantly improve the transmission efficiency and the smoothness of the user experience. Since the user's movement speed is typically on the order of seconds, and outpainting requires generating scenes with a lot of detail, DMs are suitable for outpainting.

\end{itemize}

\subsection{Content Caption Module}
The prediction accuracy challenge could be alleviated by providing more inputs. For example, text descriptions about XR content and user actions can be utilized to improve the similarity between the generated content and the original content. Therefore, a content caption module is proposed to translate XR content and user actions into text descriptions at the cloud server.

The content caption module extracts textual descriptions from video frames, including visual elements, user actions, and context, to enhance prediction accuracy. This process begins with using computer vision techniques such as object detection and action recognition to identify key visual features and user movements in each frame. These visual elements and actions are then fed into a transformer-based model. The transformer model processes the extracted features from multiple frames, capturing temporal dependencies and generating coherent textual descriptions that describe the XR content and the user’s actions in the scene. With the related text descriptions, the accuracy of the content generation module can be improved.


\subsection{DRL-based Decision Module}


As mentioned before, various decision-making challenges need to be solved for better utilizing GAI to improve XR content transmission efficiency. Since decision-making challenges can be modeled as a Markov Decision Process (MDP), we leverage the DRL model to propose a transmission mode decision module that consists of multiple agents for dynamically making different types of decisions. 

To perform multiple decision-making tasks, each task needs a specific agent. For most agents, the state set can be modeled as the available computational resources and network conditions (including the available link bandwidth, the ongoing traffic load, and the type of each content) that the agents can observe. The state set \( S \) is thus defined as:
$S = \{ B_a, C_a, L, T, U \}$ where \( B_a \) represents the available bandwidth, \( C_a \) stands for the available computational resources, \( L \) is the ongoing traffic load, and \( T \) represents the type of XR content, \( U \) represents user behaviors. The only exception is the state set of the cross-modal data generation model, which additionally needs to consider the modality of the current content (denoted by \( M_c \)) being processed.

However, the action set is designed differently for each agent, based on the specific task. Specifically, the action set \( A \) is defined for different agents as follows:

\begin{itemize}
    \item \textbf{Subsequent Frame Generation Agent}: The action set consists of the length and quality of the subsequent frames of the XR content to be generated. This can be expressed as: $A = \{ l_{\text{start}}, l_{\text{next}}, q_{\text{next}} \}$
    where \( l_{\text{start}} \) is the starting frame to be generated, \( l_{\text{next}} \) is the number of the generated frames of XR content, and \( q_{\text{next}} \) represents the quality of these frames.

    \item \textbf{Object Generation Agent}: In this case, the action set is defined as the objects that need to be generated at the edge servers. The action space is defined as: $A = \{ o_{\text{obj}} \}$ where \( o_{\text{obj}} \) represents the object that needs to be generated.

    \item \textbf{Outpainting Agent}: The action set for this model is a binary decision: whether or not outpainting should be performed at the edge servers, and it is represented by:
    $A = \{ p, n \}$ where \( p \) means outpainting, and \( n \) means no outpainting.

    \item \textbf{Cross-Modal Data Generation Agent}: The action set for this agent is designed to be the modality and quality of the data to be generated. It is denoted by: $A = \{ m, q_m \}$ where \( m \) stands for the modality (data type), and \( q_m \) refers to the quality of the generated data.
\end{itemize}

The objective function (denoted as \( J(\theta) \)) for this DRL module is designed to optimize the quality of experience (QoE), i.e., maximize cumulative reward over time, constrained by available network bandwidth and content QoS requirements. Specifically, the reward is calculated based on the resolution of the XR content, the variation in resolution, and the latency. 

The DRL-based decision module proactively activates GAI models by evaluating network conditions in advance, rather than reactively responding when transmission becomes insufficient. This approach mitigates the long inference latency by triggering GAI models before network conditions deteriorate. Furthermore, since the dimensions of the state and action sets of the DRL agents are limited, the computational overhead is minimal, enabling real-time inference.

\section{A Case Study}
Since video transmission is crucial for XR content and sensitive to network fluctuations, we focus on implementing subsequent frame generation in the proposed GAI-based framework.



\begin{figure*}[!htp]
\centering {\includegraphics[width=1\textwidth]{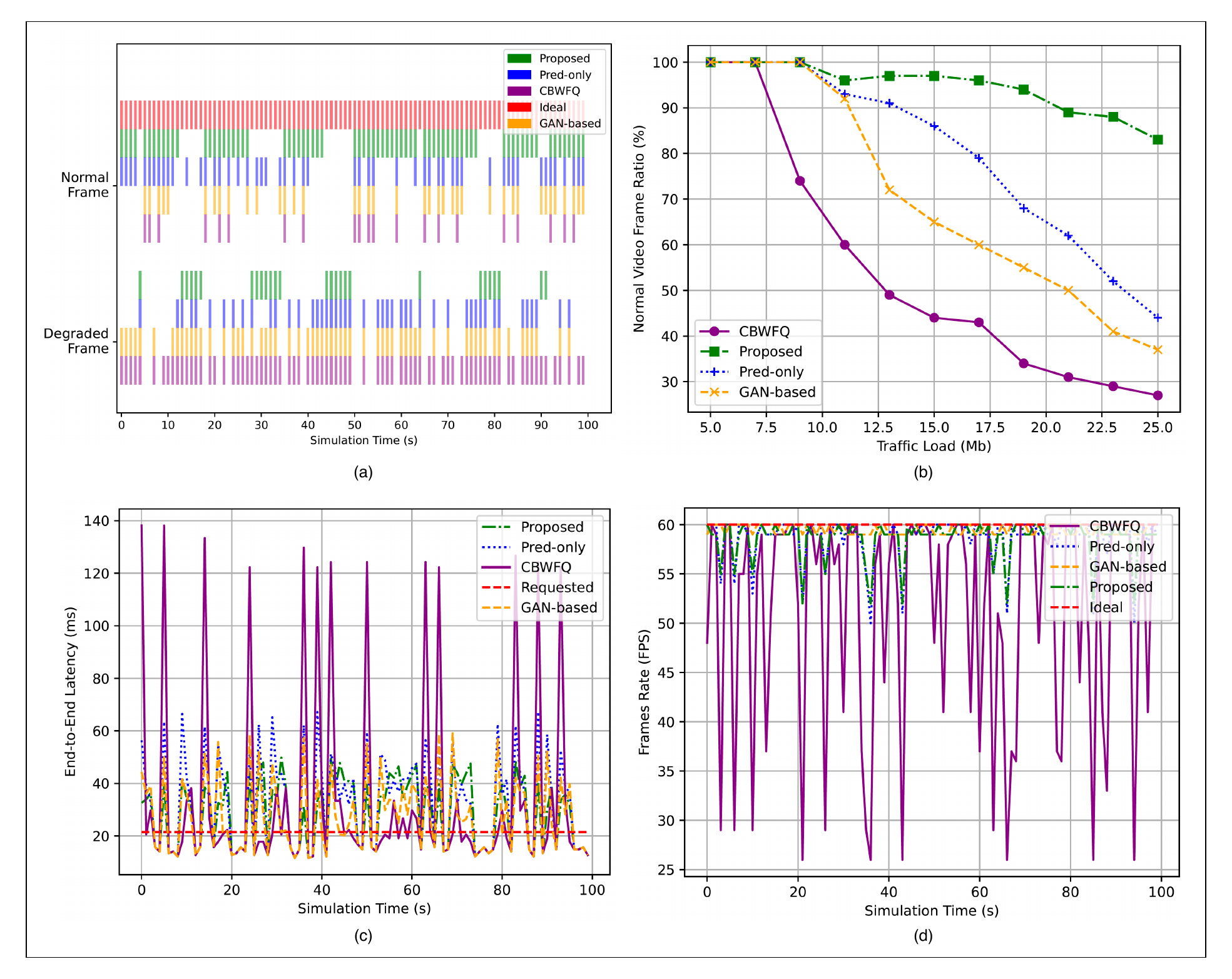}}
\caption{Performance evolution of all approaches in the simulation: (a) normal and degraded video frame distribution over the simulation time, with one frame randomly sampled per second under a fixed FPS setting (60 FPS); (b) impacts of traffic load on video frame quality; (c) the end-to-end latency over the simulation time; (d) the frame rate over the simulation time under a fixed resolution setting (2K).} 
\label{fig:main} 
\end{figure*}

\subsection{Experiment Settings}
Grand Theft Auto V (GTA V) is widely recognized as a modern prototype of the Metaverse \cite{mm} due to its ability to closely replicate real-world environments. It has been used in various research areas, such as autonomous driving. We also choose it as a representative scenario for XR content transmission in this experiment. The experiment is conducted on a Linux server equipped with dual Intel Xeon Platinum 8352V CPU, 4 RTX 4090 (24 GB) GPUs, using Python 3.9 and PyTorch 2.0. The simulator is implemented in Python, specifically designed for simulating network conditions (e.g., bandwidth, other traffic flows) and transmission of XR content. The simulation inputs include GTA V game-play frames. When network conditions fail to support the transmission of original content, the content generation module utilizes Runway's GEN-2 model (DMs-based) to generate subsequent frames. Since D3QN can provide more stable and efficient performance \cite{D3QN}, and is suitable for a wide range of real-world tasks, we adopt D3QN as the DRL model in the DRL-based decision module. 



We use the normal frame ratio, end-to-end latency, and frame rate as the evaluation metrics. The former metric is the percentage of normal frames out of all frames (normal and degraded frames). Frames are considered degraded if either the frame resolution decreases due to insufficient bandwidth allocation or if the generated frames deviate significantly from the original frames, measured by the Structural Similarity Index (SSIM) (e.g., less than 0.9). For performance comparisons, we use the following baselines:
\begin{itemize}
\item \textbf{CBWFQ}: A cloud-to-user transmission mechanism employing Class-Based Weighted Fair Queuing for scheduling.
\item \textbf{Pred-only}: An edge server-based approach that includes only the content generation module to predict subsequent XR content, highlighting the effectiveness of our DRL-based decision module.
\item \textbf{GAN-based}: A variation of Pred-only using GANs instead of DMs to evaluate sensitivity to different GAI models.
\end{itemize}


We use the network that is illustrated in Fig. \ref{proposed} as the network scenario, in which the original XR content is stored at the cloud server, and the GAI models are implemented at the edge server. The simulation represents a continuous game-play session lasting 100 s, focusing on the transmission of game content under varying network conditions. The link bandwidth is set as 26 Mbps. Since the game-play scenes are 2K 60 frames per second (FPS), the bandwidth requirement for smooth transmission is at least 16 Mbps. H.266 is used as the compression technique. To simulate the dynamic network conditions, we randomly introduce other traffic flows (5 to 25 Mb) throughout the entire simulation. 

\subsection{Experiment Results}
The performance of all approaches is shown in Fig. \ref{fig:main}. To ensure result reliability, we conducted multiple simulation runs, and each run produced consistent outcomes. Specifically, Fig. \ref{fig:main}. (b) illustrates the averaged results from all simulations, while Fig. \ref{fig:main}. (a), (c), and (d) represent outcomes from randomly selected runs due to the random nature of traffic load generation.

The major findings from these simulations are as follows:
\begin{itemize}
    \item \textbf{Higher Normal Frame Ratio}: The proposed approach consistently maintains a higher normal frame ratio of 84\%, benefiting from the DRL-based decision module and the high quality of frames generated by DMs. In contrast, the GAN-based approach, due to lower quality of frames generated by GANs, and the Pred-only approach, lacking the DRL-based decision module, both result in a greater number of degraded frames.

    \item \textbf{Adaptability to High Traffic Load}: Fig. \ref{fig:main}. (b) shows that our approach has the strongest resistance to high traffic load, it can still achieve more than 90\% normal frame ratio even when the traffic load is around 20 Mb. In contrast, baseline approaches can only obtain less than 65\% normal frame ratio. The reason behind this phenomenon is that the DRL-based decision module can dynamically adjust the quality and number of the generated frames to fit the available bandwidth.

    \item \textbf{Stable Performance with Low Latency and High FPS}: Fig. \ref{fig:main}. (c) demonstrates that the proposed approach can maintain the end-to-end latency around 37 ms with minimal fluctuations. This is a result of the proactive frame generation and transmission adjustments enabled by the DRL-based decision module. In contrast, CBWFQ fluctuates dramatically due to inefficient bandwidth. GAN-based and Pred-only approaches also underperform, sometimes even worse than CBWFQ, because they cannot adapt to network changes without the DRL-based decision module.

    As illustrated in Fig. \ref{fig:main}. (d), GAI-based approaches achieve significantly higher and more stable frame rates (exceeding 50 FPS) compared to CBWFQ, which drops below 27 FPS in the worst case. The stability of GAI-based approaches stems from their ability to generate subsequent frames when bandwidth is insufficient, unlike CBWFQ, which is highly susceptible to bandwidth fluctuations. Among these approaches, the GAN-based method delivers the best performance, as its faster frame generation compared to DMs (used in both the proposed approach and Pred-only) further enhances the frame rate.

\end{itemize}

\subsection{Discussions of the Proposed Framework}
\subsubsection{Scalability} The scalability of the proposed framework is mainly impacted by the GAI models since they require significant computational resources for inference. However, our framework only performs inference when the network fails to meet XR content transmission requirements, and it only generates the affected parts, not the entire XR content. Hence, the framework is highly scalable with moderate computational resources.
\subsubsection{Limitations} The proposed approach depends on the GAI model’s prediction accuracy; large deviations from real content could negatively impact user experience. Additionally, GAI requires significant computational resources, meaning content must be pre-generated. Quick detection or prediction of network changes is crucial for maintaining performance. Since the inference time of GAI models is at the ms level, the detection or prediction of network changes should be less than it. 

Another critical limitation of the proposed framework lies in the potential privacy risks associated with generating XR content on edge servers. Because this process may involve sensitive user data, such as the text description and first frame of the original XR content. More techniques, including encryption, access control, and anonymization, are required to address this limitation.

\subsubsection{Potential Impacts} The proposed approach addresses critical network challenges for XR content transmission, including adaptive handling of large data volumes, seamless interaction, stable performance under dynamic conditions, and ubiquitous access across diverse devices. By tackling these issues, the framework advances content transmission technologies, supporting the evolving demands of the Metaverse and fostering future innovations.


\section{Conclusion and Future Directions}

This work discussed the network challenges of XR content transmission in the Metaverse and explored how GAI can overcome them. The proposed GAI-based framework can facilitate XR content transmissions from the content generation perspective (e.g., generating subsequent frames). The results of the case study validate the effectiveness of the proposed framework. This work offers a scalable solution adaptable to various immersive applications, advancing the realization of the Metaverse over current networks. It provides valuable insights for academic and industry researchers aiming to enhance XR content delivery over current networks.

Future research should focus on minimizing the discrepancies between GAI-generated and original content. Additionally, networks need to become more intelligent for real-time, low-latency decision-making to adapt to varying traffic loads. Given the resource-intensive nature of GAI models, there is also a need to explore lightweight GAI models to reduce computational costs.


\bibliographystyle{IEEEtran}
\bibliography{reference}

\begin{IEEEbiographynophoto}{Zhe Zhang}[SM] received his Ph.D. degree in Electrical and Computer Engineering from Carleton University, Ottawa, Canada in 2019. Currently, he is an Associate Professor at Nanjing University of Posts and Telecommunications, Nanjing, Jiangsu, China. His research interests include multimedia networking, intelligent networks, Metaverse, and in-network caching.
\end{IEEEbiographynophoto}

\begin{IEEEbiographynophoto}{Yili Jiang} received her Ph.D. degree in Computer Engineering from the University of Nebraska-Lincoln, NE, USA, in 2022. She is currently an assistant professor in the Department of Computer Science at the Georgia State University, GA, USA. Her research interests include cybersecurity, machine learning, cloud/edge computing, and wireless networks.
\end{IEEEbiographynophoto}

\begin{IEEEbiographynophoto}{Xin Wei}[SM] received his Ph.D. degree major at information and communication engineering from Southeast University, Nanjing, China in 2009. Now, he is a professor in Nanjing University of Posts and Telecommunications, China. His research interests are in the areas of multimedia communications and computing, educational technology.
\end{IEEEbiographynophoto}

\begin{IEEEbiographynophoto}{Mingkai Chen}[M'19] received a Ph.D. degree in information and communication engineering from Nanjing University of Posts and Telecommunications, Nanjing, China, in 2019. He is currently an Associate Professor at Nanjing University of Posts and Telecommunications. His research interests are in the area of multimedia communications and computing, resource allocation, and signal processing in wireless networks.
\end{IEEEbiographynophoto}

\begin{IEEEbiographynophoto}{Haiwei Dong}[SM] (haiwei.dong@ieee.org) is a Director and Principal Researcher at Huawei Canada, and an Adjunct Professor at the University of Ottawa. His research interests include artificial intelligence, multimedia, Metaverse, digital twin, and robotics.
\end{IEEEbiographynophoto}


\begin{IEEEbiographynophoto}{Shui Yu}[F] (shui.yu@uts.edu.au) is a Professor with the School of Computer Science, Deputy Chair of University Research Committee, University of Technology Sydney, Australia. His research interests include cybersecurity, network science, big data, and mathematical modeling. He has published five monographs and edited two books, more than 500 technical papers.
\end{IEEEbiographynophoto}

\end{document}